\documentclass[
showpacs,preprintnumbers,amsmath,amssymb]{revtex4}
\begin{document}
\title{Thermophoresis as persistent random walk}
\author{A.V. Plyukhin}
\email{aplyukhin@anselm.edu}
 \affiliation{ Department of Mathematics,
Saint Anselm College, Manchester, New Hampshire 03102, USA 
}

\date{\today}

\begin{abstract}
In a simple model of a continuous random walk a 
particle moves in one dimension with the velocity fluctuating 
between $+v$ and $-v$. If $v$ is associated with the thermal 
velocity of a Brownian particle and allowed to be position dependent,
the model accounts readily for the particle's drift along the temperature 
gradient and 
recovers basic results of the 
conventional thermophoresis theory.
%
\end{abstract}


\maketitle

Thermophoresis~\cite{Piazza,Piazza2,Duhr,Bring} is the systematic drift of a Brownian particle 
caused by a temperature 
gradient $\nabla T$. The steady-state thermophoretic velocity acquired by 
the particle  is given by 
\begin{eqnarray}
V_T=-D_T\nabla T,
\end{eqnarray}
where  $D_T$ is generally referred to as thermal diffusion coefficient.
A similar phenomenon  exists on the  molecular level 
in gas mixtures and is known as the  Soret effect. 
In most cases $D_T$ is positive, i.e. 
the particle moves toward a colder region.
However, thermophilic behavior, $D_T<0$,  has also been observed 
experimentally~\cite{TP1,TP2,TP3}.
Among other well-known field-driven transport effects,
thermophoresis is perhaps the most subtle: 
it takes place in a stationary
state when the pressure is uniform, and therefore 
the average force on a particle which is {\it fixed in space} 
is strictly zero. 
Interesting in itself, thermophoresis has many applications, in particular 
in the context of molecular Brownian 
motors driven by thermal fluctuations~\cite{Buttiker,Landauer,Astumian,Hondou,Kawai}.

Despite many efforts, 
a realistic theoretical 
model of thermophoresis, valid over a wide region of densities of the host 
fluid,  seems to be lacking, and even the physical origin of the 
effect is still under debate~\cite{Piazza}. Many authors argue  
that the effect cannot be described by a simple modification of 
the over-damped 
diffusion equation in the configurational space only (Smoluchowski equation), 
but requires a kinetic description.
Well-known example of such approach is that by van Kampen~\cite{Kampen},
which takes as a starting point the Kramers equation with 
position-dependent temperature
\begin{eqnarray}
\frac{\partial f}{\partial t}=-v \frac{\partial f}{\partial x}
-\frac{F}{m}\frac{\partial f}{\partial v}+\gamma\frac{\partial }{\partial v}
\left(
v f+\frac{kT(x)}{m}\frac{\partial f}{\partial v}
\right).
\label{Kramers}
\end{eqnarray}
Here $f(x,v,t)$ is the 
joint probability density of position and velocity,   
$F$ is an external force, and $\gamma$ is the friction coefficient
which may depend on $x$ and is assumed to be large.
Using the expansion in powers of $\gamma^{-1}$, one can obtain 
to the lowest order the corresponding 
overdamped Smoluchowski equation for the spatial 
density $f(x,t)$ in the form
\begin{eqnarray}
\frac{\partial f}{\partial t}+\frac{\partial }{\partial x}J=0
\label{Smoluch},\,\,\,\,\,\,\,\,J=
\left(
\mu F-D\frac{\partial }{\partial x}-D_T\frac{\partial T}{\partial x}
\right) f(x,t),
\end{eqnarray}
where the mobility $\mu=1/m\gamma$, the diffusion coefficient 
$D=kT\,\mu$, and  
the thermal diffusion coefficient is
\begin{eqnarray}
D_T=\frac{k}{m\gamma}=\frac{D}{T}=k\mu.
\label{result1}
\end{eqnarray}
Therefore, the thermophoretic force $F_T$ on the particle, defined through 
the relation $\mu F_T=-D_T\nabla T$, equals
\begin{eqnarray}
F_T=-k\nabla T.
\label{result2}
\end{eqnarray} 
The result (\ref{result1}) 
often underestimates the value of $D_T$ 
by up to one order of 
magnitude~\cite{Piazza}, and cannot, of course, account for thermophilic behavior corresponding to a negative $D_T$. Another restriction of the approach is that 
the validity of the Kramers equation for the case of 
nonuniform temperature is postulated rather than proved. 
More ``microscopic'' approaches
lead to the expression for $D_T$ with additional terms involving rather 
complicated correlation functions~\cite{K1,K2,K3}.
However,  conceptually the result (\ref{result1}) is important as a 
demonstration that 
the expression for the current $J$ in 
the Smoluchowski equation (\ref{Smoluch})  
cannot be obtained by simple insertion of the position dependence into
the conventional drift-diffusion expression 
$J=(\mu F-D\partial/\partial x)f$
for an isothermal  system.

The simplicity of the result (\ref{result1}) suggests that it might 
have a transparent physical interpretation.
Yet the transition from 
the Kramers equation (\ref{Kramers}) in phase space to equation 
(\ref{Smoluch}) for the spatial distribution function $f(x,t)=\int dv f(x,v,t)$
(the elimination of the fast variable $v$) is not a trivial 
step~\cite{Titulaer,Sancho}, which makes
the origin of the result (\ref{result1}) somewhat obscure. 
The same perhaps may be said of
the Luttinger's method of fictitious external 
fields~\cite{Lut}, which 
also leads to the result (\ref{result1})~\cite{Efros,Bash}.
The aim of this Letter is to
formulate a minimal qualitative 
model of thermophoresis, which leads to the expression (\ref{result1})
elementary and directly. 
Besides  
pedagogical merits, such model might be useful for numerical modeling of
stochastic processes at nonuniform temperature. 

The model is a slightly generalized version of the 
well-known stochastic process of the continuous persistent 
random walk~\cite{Gold,Kac,Weiss}. 
In its simplest setting, the process describes a particle moving in one dimension with fixed speed $v$ suffering occasionally a complete 
reversal of direction. Let $f_+(x,t)$ and $f_-(x,t)$ 
be the probability density for the particle moving to the right and to the 
left, respectively.  Reversals of velocity are Poisson distributed, i.e. 
occurring with a constant rate $1/2\tau$, so that    
the probability for reversal in a time interval $dt$  is $dt/2\tau$.    
For an infinitesimal time step  one can write
\begin{eqnarray}
f_+(x,t+dt)=
f_+(x-v\,dt,t)-f_+(x-v\,dt, t)\,dt/2\tau+f_-(x+v\,dt,t)\,dt/2\tau,
\end{eqnarray}
and a similar equation for $f_-(x,t)$. 
The corresponding differential equations read 
\begin{eqnarray}
\frac{\partial f_+}{\partial t}=-v\frac{\partial f_+}{\partial x}
-\frac{f_+-f_-}{2\tau},\,\,\,\,\,\,\,\,\,\,
\frac{\partial f_-}{\partial t}=v\frac{\partial f_-}{\partial x}
+\frac{f_+-f_-}{2\tau},
\label{e}
\end{eqnarray} 
and  lead to the telegrapher's equation 
for the total density       
$f=f_++f_-$, 
\begin{eqnarray}
\frac{\partial^2 f}{\partial t^2}+
\frac{1}{\tau}\,\frac{\partial   f}{\partial t}=
v^2\,\frac{\partial^2 f}{\partial x^2}.
\label{TE}
\end{eqnarray}
The same equation holds also for the difference $\Delta=f_+-f_-$, and
therefore for the components $f_+$ and $f_-$  separately.

Suppose the particle at $t=0$ is  at the origin with equal probability 
to be  in each of the two velocity states, 
\begin{eqnarray}
f_\pm(x,0)=\frac{1}{2}\delta(x),\,\,\,\,\,
\frac{\partial f_\pm(x,0)}{\partial t}=\mp\frac{v}{2}\frac{\partial}{\partial x}\,\delta(x).
\end{eqnarray}
Here the second initial condition follows from the first one and 
Eqs.(\ref{e}). Respectively, the 
initial conditions for the  total density  $f=f_++f_-$ are
\begin{eqnarray}
f(x,0)=\delta(x),\,\,\,\,\,\frac{\partial f(x,0)}{\partial t}=0.
\label{BC1}
\end{eqnarray}
The corresponding solution of the telegrapher's equation is well 
known~\cite{Zau}. In the long-time limit $t\gg \tau$, $vt\gg x$ it 
coincides exactly with solution of the 
the Smoluchowski equation
\begin{eqnarray}
f(x,t)\approx
(4\pi Dt)^{-1/2}\exp\left(-\frac{x^2}{4Dt}\right)
\label{diffusion}
\end{eqnarray}
with the diffusion coefficient $D=\tau\,v^2$. 
However, unlike the overdamped Smoluchowski equation (\ref{Smoluch}),
equations (\ref{e}) incorporates effects 
of inertia of the particle. This advantage, which  was recognized and used 
beneficially in many previous works (see~\cite{Weiss} and references therein), 
allows to account for thermophoresis 
in a particularly  simple way.   

To link the model to the problem of thermophoresis, it is 
natural to identify $v$ with a typical thermal speed of a Brownian particle.
The choice is ambiguous.
For instance, one can set $v$ equal to the root mean square velocity
$v_{rms}=\langle v^2\rangle^{1/2}=\sqrt{3kT/m}$. 
However, one can show that in this
case the final result for the thermophoretic force would differ from 
Eq.(\ref{result2}) by the factor $3/2$ (see Eq.(\ref{Ft}) below). 
As will be shown, a perfect  agreement with the
standard results $D_T=D/T$ and $ F_T=-k\nabla T$ can be achieved 
if $v$ is identified not with $v_{rsm}$ but with the most 
probable speed of a Brownian particle $v_{mp}$ 
(for which the Maxwell 
speed distribution has a maximum): 
\begin{eqnarray}
v= v_{mp}=\sqrt{2kT/m}.
\end{eqnarray}
The second parameter of the model $1/\tau$ should be identified 
with the friction 
coefficient $\gamma$ which appears in the Kramers equation (\ref{Kramers})
and in the corresponding Langevin equation $\dot v=-\gamma v +\xi(t)$.

For nonuniform temperature,  the velocity in Eqs. (\ref{e})
is position dependent, $v(x)=\sqrt{2kT(x)/m}$.  
In this case instead of the 
telegrapher's equation (\ref{TE})
one obtains~\cite{Weiss2}
\begin{eqnarray}
\frac{\partial^2 f}{\partial t^2}+
\frac{1}{\tau}\,\frac{\partial   f}{\partial t}=
v^2\,\frac{\partial^2 f}{\partial x^2}+
\left(v\frac{dv}{dx}\right)
\,\frac{\partial f}{\partial x}.
\label{TE2}
\end{eqnarray} 
Suppose the temperature gradient is constant, so that $T(x)=T+x\,\nabla T$ and
\begin{equation}
v(x)=\sqrt{2kT(x)/m}=v\left(1+\frac{\nabla T}{T}x\right)^{1/2},
\end{equation} 
where $v$ is the thermal velocity corresponding to the temperature $T$,
$v=\sqrt{2kT/m}$. 
Then the  equation (\ref{TE2}) reads as  
\begin{eqnarray}
\frac{\partial^2 f}{\partial t^2}+
\frac{1}{\tau}\,\frac{\partial   f}{\partial t}=
v^2\left(1+\frac{\nabla T}{T}x\right)
\,\frac{\partial^2 f}{\partial x^2}
+v^2\,\frac{\nabla T}{2T}\,
\,\frac{\partial f}{\partial x}.\nonumber
\label{TE3}
\end{eqnarray}  
For a small gradient $(\nabla T/T)x\ll 1$,
the equation is simplified to the form  
\begin{eqnarray}
\frac{\partial^2 f}{\partial t^2}+
\frac{1}{\tau}\,\frac{\partial   f}{\partial t}=
v^2\,\frac{\partial^2 f}{\partial x^2}
-\frac{F_T}{m}\,
\,\frac{\partial f}{\partial x},
\label{TE4}
\end{eqnarray} 
where  the thermophoretic force $F_T$ coincides (thanks to the setting
$v=\sqrt{2kT/m}$) with 
the expression   (\ref{result2}) of the standard theory: 
\begin{eqnarray}
F_T=-\frac{1}{2}mv^2\,\frac{\nabla T}{T}=-k\,\nabla T.
\label{Ft}
\end{eqnarray}
It can  be shown that in the long-time limit 
the solution of the equation (\ref{TE4}) with the 
boundary conditions (\ref{BC1}) 
coincides with the solution of the
overdamped equation (\ref{Smoluch}) and  
describes the drift along  
the temperature gradient, superimposed on the diffusion 
\begin{eqnarray}
f(x,t)\approx\frac{1}{\sqrt{4\pi D t}}\,\exp\left(
-\frac{(x-V_Tt)^2}{4Dt}
\right)
\label{result3}
\end{eqnarray}
with the drift velocity $V_T=F_T\tau/m$.
Using  (\ref{Ft}) and recalling $D=\tau v^2$, one gets 
$V_T=-D_T\nabla T$ with $D_T=D/T$, thus recovering the result 
(\ref{result1}) of the conventional theory.

The asymptotic solution (\ref{result3}) 
can be obtained as follows. After applying the 
transformation $f(x,t)=\phi(x,t)\,\exp(-t/2\tau+F_Tx/2mv^2)$, the equation 
(\ref{TE4}) takes the form
\begin{eqnarray}
&&\frac{\partial^2 \phi}{\partial t^2}=
v^2\,\frac{\partial^2 \phi}{\partial x^2}+\frac{1}{\tau_*^2}\phi,
\label{TE5}
\end{eqnarray}  
with $1/\tau_*^2=1/4\tau^2-F_T/4m^2v^2$, 
while the boundary conditions corresponding to (\ref{BC1}) read 
\begin{eqnarray}
\phi(x,0)=\delta(x),\,\,\,\,\,
\frac{\partial \phi(x,0)}{\partial t}=\frac{1}{2\tau}\delta(x).\nonumber
\label{BC2}
\end{eqnarray}  
Note that the model makes sense only 
under the assumption $\tau_*^2>0$, which guarantees the positiveness of $f_+$ 
and $f_-$~\cite{Kim}. The equation (\ref{TE5}) is 
the modified telegrapher's equation whose solution is well known~\cite{Zau}. 
Transforming back from $\phi$ to $f$, the result can be written in the following 
form
\begin{eqnarray}
f(x,t)=\exp\left(
-\frac{t}{2\tau}+\frac{F_Tx}{2mv^2}
\right)\,(\phi_1+\phi_2+\phi_3),
\label{exact}
\end{eqnarray}
where the functions $\phi_i(x,t)$ are
\begin{eqnarray}
\phi_1(x,t)&=&\frac{1}{2}[\delta(x-vt)+\delta(x+vt)],\\
\phi_2(x,t)&=&\frac{1}{4v\tau}\,I_0\left(
\frac{1}{v\tau_*}\sqrt{v^2t^2-x^2}\right)\,
\theta(vt-|x|),\nonumber\\
\phi_3(x,t)&=&\frac{t}{2\tau_*\sqrt{v^2t^2-x^2}}\,I_1\left(
\frac{1}{v\tau_*}\sqrt{v^2t^2-x^2}\right)\,
\theta(vt-|x|),\nonumber
\end{eqnarray}
and $\theta(x)$ is the unit step function. 
Using the  
asymptotic form of the 
modified  
Bessel functions for large argument 
$I_\alpha(x)\approx\frac{1}{\sqrt{2\pi x}}e^x$, 
it is an easy matter to prove that 
in the limit of strong damping
$F_T\tau/mv\ll 1$
and of long time $t\gg\tau$,
$vt\gg x$, 
the exact solution (\ref{exact}) is reduced to the simple 
form (\ref{result3}). 

One interesting problem related to
thermophoresis is that of Brownian motors driven by position-dependent 
temperature. 
In particular, the  B\"uttiker-Landauer motor~\cite{Buttiker,Landauer} 
is essentially a 
Brownian particle diffusing in the periodic potential field 
subject to a spatially 
inhomogeneous temperature.  
In this context it is of interest to generalize
the model to the case of  the presence of an external force $F$.
As discussed in~\cite{Kim}, in this case the equations (\ref{e}) for 
$f_\pm(x,t)$ should be extended as follows 
\begin{eqnarray}
\frac{\partial f_\pm}{\partial t}=\mp v\,\frac{\partial f_\pm}{\partial x}
\mp\frac{f_+-f_-}{2\tau}\pm\frac{F}{2mv}\,(f_++f_-).
\label{ee1}
\end{eqnarray}    
Respectively, the equation (\ref{TE4})
for the total density $f=f_++f_-$ is generalized to the form 
\begin{eqnarray}
\frac{\partial^2 f}{\partial t^2}+
\frac{1}{\tau}\,\frac{\partial   f}{\partial t}=
v^2\,\frac{\partial^2 f}{\partial x^2}
-\frac{F_T}{m}\,
\,\frac{\partial f}{\partial x}-\,
\frac{\partial }{\partial x}
\left(
\frac{F}{m}f
\right).
\label{ee2}
\end{eqnarray}
This equation differs from the overdamped equation (\ref{Smoluch}) 
of the standard theory by the presence of the term $f_{tt}$.
This term, related to inertial effects,  is unimportant 
in the long time limit, 
but may be responsible for
wave-like behavior at short times~\cite{Weiss}.

Summarizing, in this Letter we discussed a stochastic process which 
underlies thermophoresis in a way similar to that as the discrete 
random walk underlies isothermal diffusion. The model leads to the
telegrapher's equation whose asymptotic solution 
coincides with the solution of 
the overdamped Smoluchowski equation (\ref{Smoluch}). 
This is consistent with an observation that the telegrapher's equation
is reduced to the  Smoluchowski
equation under conditions
$v\to\infty$, $\tau\to 0$, $v^2\tau=const$.
Since the model takes into account inertial effects, one might hope that
it can 
resolve difficulties
of the overdamped theory of thermally driven 
Brownian motors~\cite{Astumian,Hondou,Kawai}.
However, similar to the Kramers
equation (\ref{Kramers}),   the model
is based on the assumption that
the thermalization of the particle to a local temperature is 
faster than any other process involved. Since the characteristic time
scales for inertial and thermalization effects are typically the same, 
the application 
of the model beyond the overdamped regime, strictly speaking, is  
not justified,
and should be undertaken with caution. 
Despite of this limitation, the mapping of 
thermophoresis onto a random walk problem may offer
some advantage for both analytical and numerical modeling, 
in particular for problems with complicated or
velocity-dependent~\cite{Weiss3,Kim}  boundary conditions.



\end{document}